\documentclass[11pt,letterpaper]{article}
\usepackage{arxiv}
\usepackage[pdftex]{graphicx}
\usepackage{graphicx}
\usepackage{rotate}
\usepackage{wrapfig}
\usepackage{caption}
\usepackage[T1]{fontenc}
\usepackage[english]{babel}
\usepackage{times}
\usepackage{amsmath,amstext,amsbsy,amsopn,amsthm,amsfonts,amssymb}
\usepackage{mathtools}
\usepackage{url}
\usepackage{tabularx,booktabs}
\newcolumntype{C}{>{\centering\arraybackslash}X}
\usepackage{booktabs,array,colortbl}
\usepackage{lastpage}
\usepackage{multirow}
\usepackage{algorithm}
\usepackage{balance}
\usepackage{placeins}
\usepackage{array}
\usepackage{blindtext}
\usepackage{subfigure}
\usepackage{caption}
\usepackage{capt-of}
\usepackage{booktabs}
\usepackage[small]{titlesec}
\usepackage{paralist}
\usepackage{verbatim} 
\usepackage[usenames,dvipsnames,svgnames]{xcolor}
\usepackage{algorithmic}
\usepackage{cite}
\usepackage{fancyhdr}
\usepackage{wrapfig}
\begin{document}

\title{Smart Healthcare for Diabetes: \\ A COVID-19 Perspective}

\author{
%
\begin{tabular}{ccc}
	\\
	Amit M. Joshi & Urvashi P. Shukla & Saraju P. Mohanty \\
Dept. of ECE &  Dept. of ECE  & Computer Science and Engineering \\
	MNIT, Jaipur, India. &  MNIT, Jaipur, India. & University of North Texas, USA\\
	Email: amjoshi.ece@mnit.ac.in & shuklaurvashiec50@yahoo.com   &
	Email: saraju.mohanty@unt.edu
	\\
\end{tabular}
}

\maketitle

\begin{abstract}
Diabetes is considered as an critical comorbidity linked with the latest coronavirus disease 2019 (COVID-19) which spreads through Severe Acute Respiratory Syndrome Coronavirus 2 (SARS-Cov-2). The diabetic patients have higher threat of infection from novel corona virus. Depending on the region in the globe, 20\% to 50\% of patients infected with COVID-19 pandemic had diabetes. The current article discussed the risk associated with diabetic patients and also recommendation for controlling diabetes during this pandemic situation. The article also discusses the case study of COVID-19 at various regions around the globe and the preventive actions taken by various countries  to control the effect from the virus. The article presents several smart healthcare solutions for the diabetes patients to have glucose insulin control for the protection against COVID-19.
\end{abstract}

\keywords{Smart Healthcare, Healthcare Cyber-Physical System (H-CPS), Internet-of-Medical-Things (IoMT), Glucose Level, Diabetes, Coronavirus, Glucose Insulin Control, COVID-19, SARS-CoV-2}

\section{Introduction}

Diabetes occurs when the body of a person finds the difficulty to balance glucose level during various prandial states \cite{Yin_TETC_2019-2958946}. Diabetes cases has exponentially increased in the past decades around the world \cite{Zhang2011, Jain_WF-IoT_2020_iGLU}. The unhealthy lifestyle is the prominent factor in magnifying the chance of being a diabetic patient. The unbalanced diet is one of the main factors for the occurrence of diabetes Mellitus \cite{balakrishnan2011review}. An estimated 463 million adults worldwide have diabetes and addressing their quality of life through smart healthcare technologies can have significant social impact \cite{diabetesatlas_URL_2020, Joshi_ISVLSI_2020_Secure-iGLU}. The main cause of the diabetes is deficiency of insulin level in the body against the generated glucose. The diabetes problem may lead to the reduction of blood pressure and other cardiovascular disease.

\begin{figure}[htbp]
	\centering
	\includegraphics[width=0.80\textwidth]{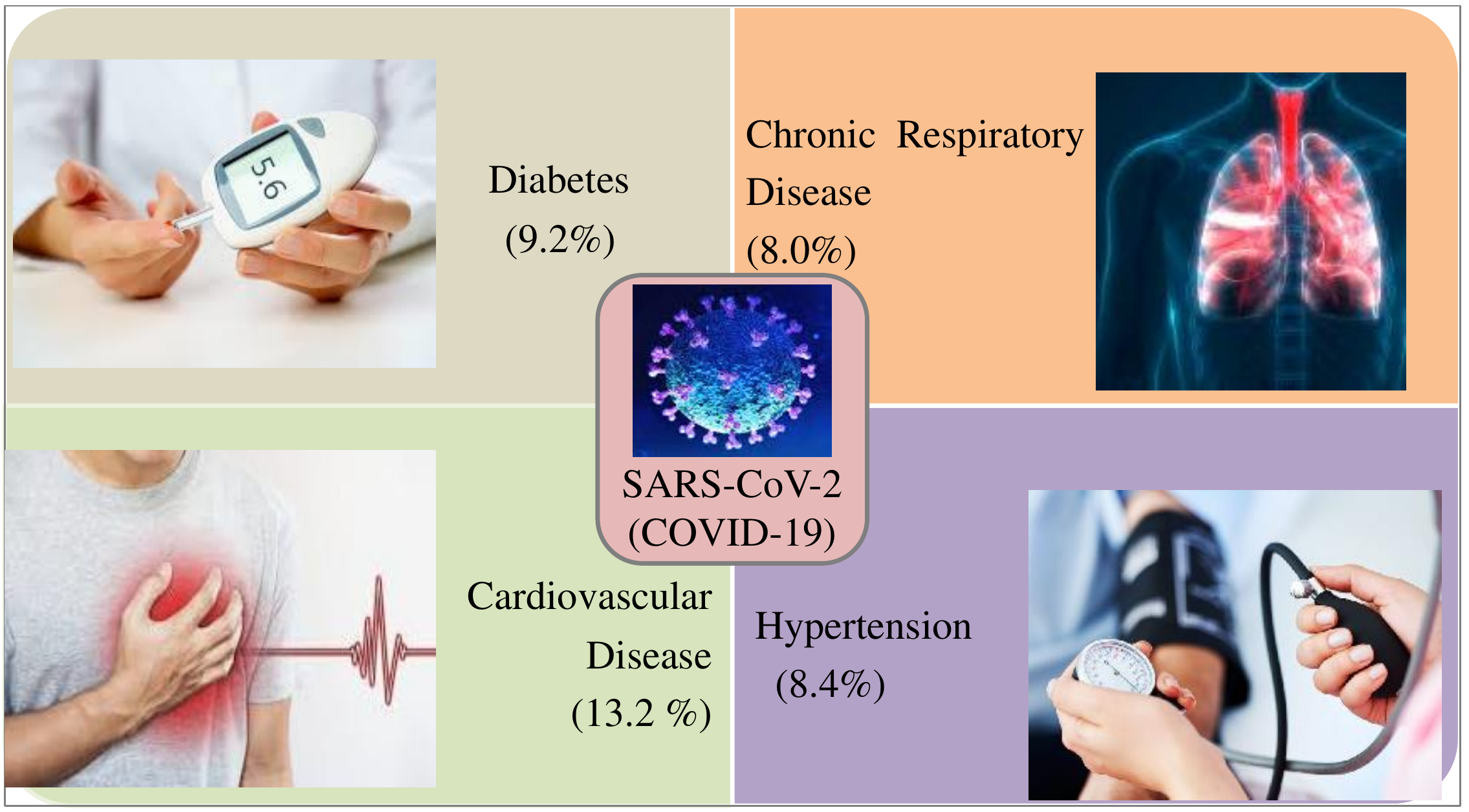}
	\caption{Comorbidities with Pre-existing medical conditions for COVID-19.}
	\label{FIG:Comorbidities_of_Pre-existing_Medical_Conditions}
\end{figure}

Severe acute respiratory syndrome coronavirus 2 (SARS-CoV-2) has infected millions of people around the globe leading to a large number of deaths \cite{World:Corona}. The Corona Virus Disease (COVID-19) has been spreading among the people worldwide at an exponential rate. SARS-CoV2 affects population of various age groups and person with underneath health conditions. The relation of COVID-19 with  underneath disease is as shown in Fig. \ref{FIG:Comorbidities_of_Pre-existing_Medical_Conditions}.

The diabetes was associated as the major risk for the past severe disease such as Severe Acute Respiratory Syndrome (SARS), Middle East respiratory syndrome (MERS) types of infections from the corona virus and the severe influenza A H1N1 pandemic in 2009 \cite{HUSSAIN2020108142}.
The diabetes persons are in bigger threat for the infection of the novel corona virus disease (COVID-19) \cite{gupta2020diabetes}. In general, the mortality rate of diabetes patients are 2 to 3 times more than other and the chances of intensive care is also higher than non diabetic patients. It is observed that the diabetes patients area at  higher risk (up to 50 \%) from the virus \cite{dong2020interactive}.

The studies show higher mortality rates of COVID-19 patients with pre-existing diabetes condition. It is difficult to provide the treatment for diabetic patients due to fluctuating of blood glucose levels. The reasons for diabetic patients serious infections by COVID-19 include the following \cite{Michael020diaCOVID-19}: 
\begin{enumerate}
	\item 
The high blood sugar would affect the immunity of the patient which made him vulnerable against the Corona virus infection as well lead to longer recovery period.

\item
The virus would thrive in an environment of the increase blood glucose. 
\end{enumerate}

The rest of the article is organized in the following manner: Section \ref{Sec:Why_Are_High-Risk} discusses the reasons for high risk for diabetic patients. Section \ref{Sec:Challenges_for_Diabetic_Patients} discusses the challenges for diabetic patients during COVID-19. A summary of various case studies for COVID-19 with respect to diabetic patients is provided in Section \ref{Sec:Case_Study_of_Diabetic_Patients}.  Section \ref{Sec:Diabetes_Management_during_COVID-19} presents some thought on diabetes management during pandemic outbreak to improve quality of life. We discuss the potential roles of selected emerging technologies in Section \ref{Sec:Roles_of_Emerging_Technologies}. Section \ref{Sec:Smart_Healthcare_for_Diabetes} presents some solutions for the diabetic population during pandemic outbreak like COVID-19. We present concluding thoughts in Section \ref{Sec:Conclusion}.

\section{Why Are Diabetes Patients at High Risk}
\label{Sec:Why_Are_High-Risk}

\begin{figure}[htbp]
	\centering
	\includegraphics[width=0.90\textwidth]{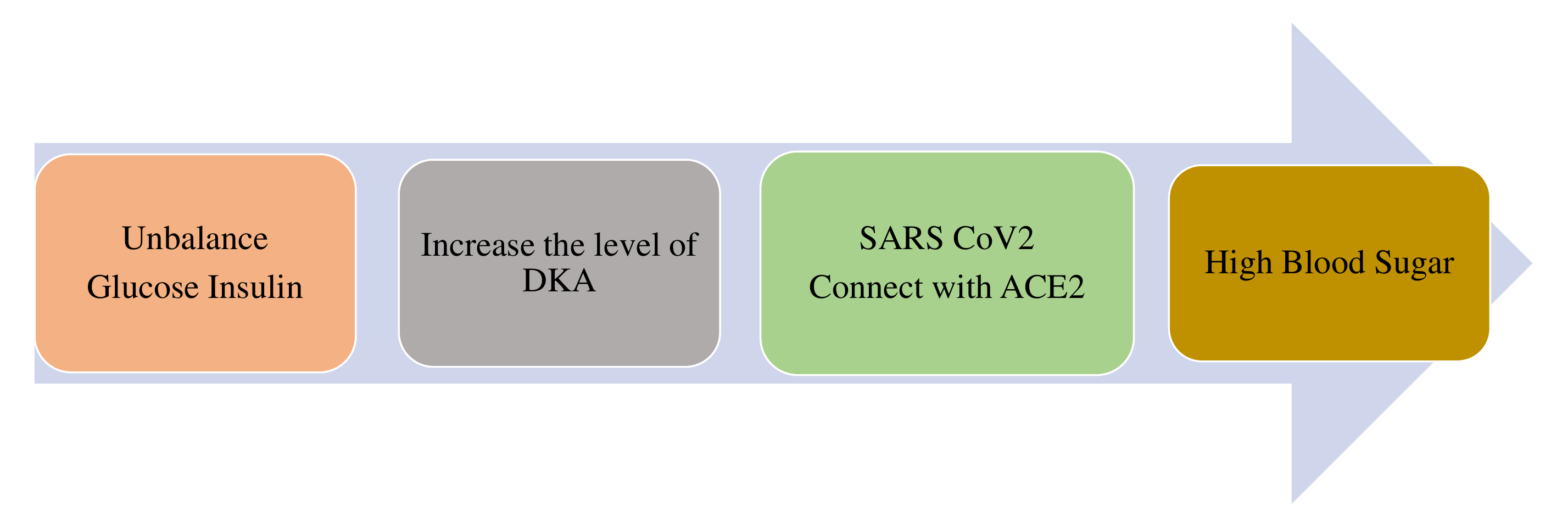}
	\caption{Impact of COVID-19 on Diabetes Patients.}
	\label{FIG:COVID-19_Impact_on_Diabetic_Patients}
\end{figure}

COVID-19 has fostered various challenges for the most vulnerable group of society ``diabetic patients'' \cite{huang2020diabetes}. Human body with underlying infection and high temperature inhabits the normal production of insulin. The role of insulin is undoubted in controlling the level of glucose in the blood. This results in a serious diabetes complication known as Diabetic Ketoacidosis (DKA). Here, the body cell don't get the glucose needed for energy, whereby the body begins to burn down the body fat for the same which results in production of excess blood acids (known as ketones) \cite{pal2020covid}. The impact of virus on the diabetes patients is shown in Fig. \ref{FIG:COVID-19_Impact_on_Diabetic_Patients}.

Studies have showed that Angiotensin-Converting Enzyme 2 (ACE2) acts as a cellular receptor for SARS-CoV-2 having 10 to 20 times higher binding affinity than the previous existing SRAS-CoV. 
ACE2 is a natural occurring enzyme present in the cell membrane of lung and enterocytes of small intestine. This duo forms a bond with their protein along with a sugar molecule. This dependency of virus leads to hyperglycemia wherein a damage is caused to pancreas islets (known as the islet cell of the pancreas) which plays a prominent role for insulin generation for blood sugar control \cite{muniyappa2020covid}.

Hyperglycemia is a characteristic of diabetes and when this chronic metabolic disorder is left untreated causes various other serious complication like kidney failure, cardio disorders and respiratory complications. This hyperglycemia environment increases the virulence of some pathogens make the patients at high risk. It is reported that phagocytes and chenotaxis are also impaired. Overall immune response is hampered which plays a vital role in fighting COVID-19. Therefore, a good monitoring over the glucose blood levels in patients with COVID-19 is essential \cite{iacobellis2020covid}.

\section{Challenges for Diabetic Patients during COVID-19 Pandemic}
\label{Sec:Challenges_for_Diabetic_Patients}

The current epidemic has influenced psychological well being of people around the world \cite{Velikic_MCE.2020.3002521, Mohanty_MCE_9122028}. It has massively affected  mental state of the people. With thoughts like anxiety of being infected by the virus, trauma of locked down and being away from the loved ones over a long period. At the same time, depression have crept in due to news of social damage done by COVID-19 around the globe. Moreover, diabetics people already suffers from anxiety and various forms of negative emotions that naturally gets intensified amid epidemic situation  \cite{cuschieri2020covid}. The non well being state of mind would somehow has been reflected in unbalanced glycemic control of diabetes people during this COVID-19 period. Therefore continuous counselling is required for self management of blood glucose control during the ongoing pandemic through collaboration with endocrinologists, psychiatrists, physician, nutritionists and diabetes educators. However, the diabetes patients have to take few self-measures and take appropriate consulting through teleconsultation mode for close coordination with physicians ensuring proper care during COVID-19 pandemic.

\subsection{Diabetes Self-Management Amid COVID-19 Pandemic}

The COVID-19 has come out as global pandemic which has affected millions of lives around the globe  \cite{shenoy2020diabetes}. The diabetes mellitus (DM) has appeared as serious comorbidity factor for increased mortality of infected people. Nevertheless, DM has played major role towards the invasive ventilation and/or intensive care unit of COVID-19 infected case  \cite{nachimuthu2020coping}. The lock down and imposed restriction on the movement by higher authority has made difficult condition for diabetic patients to control their glycemic profile.

\subsection{Diabetes Self-Management: Barriers and Solutions}

The various challenges faced by diabetes care during COVID-19 include the following:

\begin{itemize}
\item  
\textit{Lack of Confidence in Self-care}: 
It has been difficult choice to have belief on self-care device for  diabetes management. A proper counselling is required that could address to each and individual's requirement.

\item 
\textit{Technology Awareness}: 
The patient should be aware of information and communication technologies (ICT) for the use of e-health, mhealth and  telemedicine technologies which are clubbed under a big theme of Internet-of-Medical-Things (IoMT) leading to healthcare Cyber-Physical System (H-CPS) \cite{Joshi_ISVLSI_2020_Secure-iGLU}. There should be good scientific materials available in the form of books or video to guide them properly.

\item 
\textit{Proper Diet Plan}: 
The usage of high carbohydrate and saturated fats should be avoided in their regular meal. One has to set their own diet plan to maintain their calorie goals.

\item 
\textit{Economic and Social Obstruction}: 
The proper actions from the Government to ensure cost-effective therapeutic materials and provision of essential medical care should be main focus from the country government. 

\item 
\textit{Legal Barrier}: 
The healthcare providers have been always sceptical for the remote monitoring and telemedicine due to legal barrier in many counties like India. Many Government agencies have issued guidelines to enable teleconsultations service for the chronicle diseases such as diabetes.

\end{itemize}

\section{Case Study of Diabetic Patients in COVID-19 Around the World}
\label{Sec:Case_Study_of_Diabetic_Patients}

\subsection{In Asia}

India has an estimated 77 million people with diabetes \cite{saeedi2019global}. There are several reasons to have better control of diabetes during the pandemic times as compared to normal situations. First, doing so will almost certainly improve the outcomes in case someone does contract the COVID-19. All diabetes is not the same. Poorly controlled diabetes in an elderly person with heart and kidney issues is very different from well-controlled diabetes in a fit, active 40-year-old. The outcome of COVID-19 in the latter's case could well be compared to those infected but without diabetes. The seriousness of COVID-19 infection is very likely to get attenuated with better control of diabetes  \cite{mukherjee2020covid}. This itself is enough reason to ensure good sugar readings.

The second reason why diabetes control is important is the usual one - It is necessary to have a fix on our numbers (sugar, blood pressure, cholesterol) to prevent long-term complications  \cite{jain2019precise}. With population 1.3 billion, it is very difficult to maintain social distancing and also stay in  lockdown state (or stay-at-home) for a longer period. It is really difficult  to take  precautions like social distancing and staying home for the prevention of COVID-19 spread, specially for diabetes patients. The diabetic people are at higher risk of COVID-19 infection, and have bigger threat in-terms of mortality as well as morbidity. The extra precaution are required for such people to avoid any exposure to outside world.

A study of patients at China showed that diabetes patients had higher risk in terms of mortality as 7.3\% in comparison of overall rate of 2.3\% \cite{wu2020characteristics}.
Another study in China shows that the infected people of COVID-19 with diabetes reported worse results in comparison with the gender- and age-matched patients without diabetes \cite{shi2020clinical}. The different analysis from overall six study from China have observed that out of overall COVID-19 patients the average 9.7\% (6.9\%-12.5\%) patients were pre-existing diabetic patients. There was separate meta analysis was carried out on thousands COVID-19 infected people where around 8\% had historical background of diabetes. The main observation was made that diabetic people of COVID-19 infection have more mortality rate.

Russia has thousands of COVID-19 infected people along with the USA, Brazil and India.  However, it has the lower death rates. A report suggested that in Russia, the diabetes was observed one of main chronic disease along with heart disease and chronic obstructive pulmonary disease among the corona virus victims who are in intensive care \cite{Russia:Corona}.

\subsection{In Australia}

The threat for COVID-19 pandemic has been addressed in Australia \cite{Abbas_MCE_2020_3002490}.
The products for diabetic people have been provided along with insulin supply and important medicines under National Diabetes Services Scheme (NDSS) to reduce the risk of infection.
The people with underneath diabetes have suggested to obtain their regular medicines. 
Australia has already planned against COVID-19 risk for medical support with tele-healthcare, special assistance for old-age people, home delivery of the medication and better service of remotely located people.

The tele-healthcare would be provided by pharmacies to the participant who seeks any service via call. The consultancy service for telehealth would be provided to all the needed people (pregnant women, aged people over 65 and parents of new born child) for chronic disease like diabetes.
The government Australia provides assistance to diabetic patient for their safety and health related issues. The government has created one web page for the health related guidelines which is updated regularly \cite{ads2020}.

New Zealand has announced some specific guidelines for diabetes patients \cite{NZCOVID19}. The diabetic people can consult their personal doctor or general practitioner for their individual recommendation during the COVID-19 situation.  It is advisable for diabetes people to have flu vaccination to reduce the any risk for co-infection. It is recommended to maintain the glycaemic control for minimizing the threat of COVID-19 contraction. The age has also been considered as additional factor for corona virus risk.

\subsection{In Europe}

It is estimated over 59 millions people being diabetic across 44 countries and territories  \cite{godbole2020reducing}. A study shows that out of total death in UK due to COVID-19, approximately 26 \% suffered from either type-1 or type-2 diabetes  \cite{banerjee2020clinical}. Thus, making recovery of these patients a challenging task.

Italy is one of the most affected countries from the COVID-19 pandemic \cite{remuzzi2020covid}. As per a study of the hospitalized patients at University Hospital of Padova suffering from COVID-19, the diabetic people had a 8.9\% prevalence rate. Thus, the recovery of such type of people was really a challenging task.

The study carried out in Finland have evidently shown that prevention can certainly control the infection to a extend in people being overweight and suffering from glucose metabolism \cite{kumar2020normalized}. It also revealed that  higher threat among diabetic people could able to reduce upto 58\% with their daily physical activity and by proper managing their diet.

The preventive actions of diabetes could be easily applicable to other non communicable disease like (Cancer and Chronic respiratory) which are also at equal risk with diabetes. There are over 60 million diabetes patients in European countries \cite{EuropeDiabetes}. Out of which about 10.3\% of those are male and around 9.6\% are female of 25 years and older. However, the rate of prevalence has been increased from past several years among all age people which are at greater risk in this COVID-19 epidemic. The studies revealed that daily life style greatly accompanied  with unhealthy diet, obesity, overweight and lesser physical movement. These factors have even made difficult to easily recover from corona virus.

Spain is also one of the most affected country in Europe from COVID-19 along with Italy. As per a study, few thousand deaths were reported in Spain whereas the rate of prevalence of diabetes was around 12\% \cite{stoian2020diabetes}. As per same report, In Romania, approximately 50\% of people who died in COVID-19 were suffering either cardiovascular diseases or diabetes mellitus \cite{stoian2020diabetes}.

\begin{table}[htbp]
	\caption{Pre-existing Health Condition of hospitalised ICU patients in COVID-19.}
	\label{Pre-existing}
	\centering
	\begin{tabular}{|p{10.2em} | p{2.2em} | p{2.2em} | p{2.5em} | p{2.2em}|}
		\hline
		\hline
		Underlying Condition & Italy (\%)  & USA (\%) & Sweden (\%)  &  Spain (\%)  \\
		\hline		\hline		
		Diabetes & 17   & 32  & 23  &  17 \\
		\hline		
		Hypertension & 49  & NA & 34  &  NA \\
		\hline	
		Chronic lung Disease & 4   & 21  & 16  &   6 \\
		\hline	
		Cardiovascular Disease & 21  & 23  & 11  &   30 \\
		\hline	
		\hline
	\end{tabular}
\end{table}

\subsection{In North America}

In USA, more than 34.2 million people are suffering either diabetes or pre-diabetes including 14.3 millions senior citizen as per report in 2018  \cite{centers2020national}.  Elderly patients having pre-existing health condition such as diabetes, heart diseases and chronic lung disease have been reported. It was observed that 32 \% patients in Intensive Care Unit (ICU) due to COVID-19 had diabetes whereas 24 \% hospitalized patients had diabetes \cite{covid2020preliminary}. Moreover, only 6\% diabetes patients who were COVID-19 infection did not require any kind medical attention. There is no conclusive evidence about the fact that whether type-1 or type-2 diabetes patients are at more risk for COVID-19. The improper blood sugar management is more so ever reason for COVID-19 infection.

People living with diabetes require uninterrupted access to essential medicines, supplies, technologies, and care. Without insulin, a person with type-1 diabetes can potentially fatal health issues in a few days of time, and the lack, or irregular supply, of other diabetes medication, supplies, devices and/or technologies will also adversely affect the ability of all people with diabetes to manage their diabetes optimally and prevent the development of potentially fatal, short- and long-term complications.

\subsection{In South America}

Brazil is one of the most affected countries in the world due to COVID-19 infection. As per report of 2019, there are approximately 16.8 million (around 11.4\%) of people in Brazil have Diabetes between age 20 to 79 years, which is the fourth highest in the world  \cite{pititto2020diabetes}. Diabetes has been identified as one of the main threat along with other Non Communicable Disease (NCD) by SARS-CoV-2. There are approximately 71.2\% of diabetes patients in Brazil  suffering for hyperglycemia which would put them in higher threat of infection from COVID-19 \cite{barone2020impact}.

\section{Diabetes Management during COVID-19}
\label{Sec:Diabetes_Management_during_COVID-19}


A continuous monitoring of blood glucose level in diabetic patient is required \cite{Jain_IEEE-MCE_2020-Jan_iGLU1, Jain_arXiv_2020-Jan-28-2001-09182_iGLU2} for proper diagnosis. The person infected with virus and having underlying condition of diabetes will find it difficulty in balancing their glycemia profile during the aliment \cite{singh2020diabetes}. Proper insulin dosage becomes essential to mange the glucose level in order to protect against COVID-19 infection. With this vision many countries have presented certain guidelines about the self-management for diabetes patients.

Moreover, due to pandemic around certain self-care practices need to be followed by the homebound diabetes patients who are unable to carry out their routine visits \cite{bornstein2020practical}. One has to be extremely careful with the old ages diabetes patients during this COVID-19 crisis who are suffering from comorbidities as renal diseases, pulmonary, cardiovascular and  kidney. However, management of diabetes would be a real challenge in such cases and further precaution should be taken for such genre of people \cite{kumar2020diabetes}. The people with all types (type-1, type-2 and gestational), with some illness are more prone towards COVID-19.

There are several recommendations formulated for diabetic patients during COVID-19 by various  as follows \cite{rayman2020guidelines, kyrou2020covid}: 
\begin{itemize}
	\item  
	One has to drink substantial volume of fluid in order to avoid any sort of dehydration. 
	
	\item 
	The glycemic profile has to be maintained by everyone to the target value suggested by concerned doctor. Especially, the female patient with Gestational Diabetes needs a continuous measurement of glucose level that could be beneficial to control the glucose profiles  \cite{jain2019iomt}.
	
	\item 
	Continuous monitoring of blood glucose level throughout the day inhibits ketoacidosis and hypoglycemic condition. In case of type-1 diabetes patients if the value goes beyond 180 mg/dl then it is recommended to pump insulin into the blood for maintaining the glucose level.
	
	\item  
	Personal and surrounding hygienic are inevitable in any circumstances. Specifically, repeated hand washing and proper cleaning all medical equipment's such as glucometer and insulin pump with alcohol based sanitiser and soap water are important.
	
\end{itemize}

\begin{figure}[htbp]
	\centering
	\includegraphics[width=0.80\textwidth]{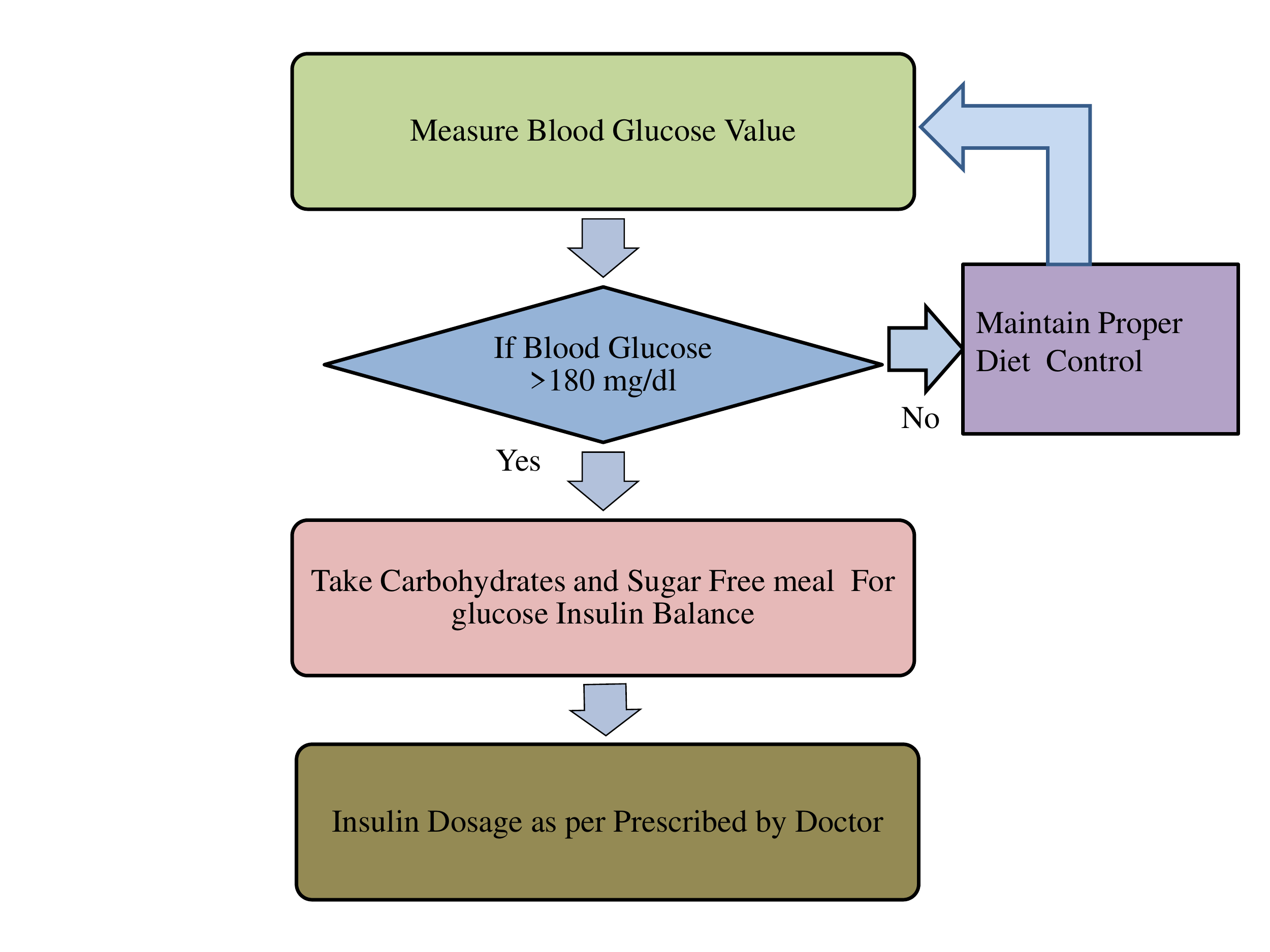}
	\caption{Balance of Glycemic Profile for Diabetes Patients during COVID-19.}
	\label{FIG:Glycemic_Profile_Balancing_Flow}
\end{figure}

It is a fact that certain medicine dose could play a crucial role in glucose control \cite{prasad2020novel}. This medicine may work towards blocking the binding of sugars and proteins. Thereby, high blood sugar value is avoided. This would theoretically impede the virus from interacting with its receptor and modulate the inflammatory response to the virus. Pre-diabetes patients are also requested to have glycated haemoglobin (HbA1c) test which calculates the average glucose level over period of 3 months  \cite{singh2020vitamin}.

High blood glucose keeps body in a low-inflammation state, thus recovery process in diabetics patients tends to be slower as compared to others. Therefore, the recommendation enlisted above are most important for diabetes people and for those who are in close contact with such people.
The balanced glycemic profile is imperative as it would help to improve the immunity \cite{ma2020covid}.

Diabetes patient has to take precautions in order to avoid major impact of virus by monitoring the blood glucose at regular interval at home and scheduled proper diet plan accordingly \cite{huang2020diabetes}. The good control of glycemic profile would help to boost the immune system of vulnerable population against COVID-19 \cite{hill2020commentary} (see Fig. \ref{FIG:Glycemic_Profile_Balancing_Flow}). The insulin secretion would help in developing immune system with balancing glucose by reducing risk of infection against the virus. The poor control of glycaemic profile may result in serious issues and also make treatment difficult for the patient. Glycaemic control of the infected person may lead to hyperglycaemia stage which makes hard to cure in presence of high fever and abnormal respiratory issues. Thus, there is a need of continuous monitoring of glucose and taking antidiabetic medication.

\section{The Roles of Emerging Technologies during Pandemic Outbreak}
\label{Sec:Roles_of_Emerging_Technologies}

\subsection{Role of Robotics}

\begin{figure}[htbp]
	\centering
	\includegraphics[width=0.88\textwidth]{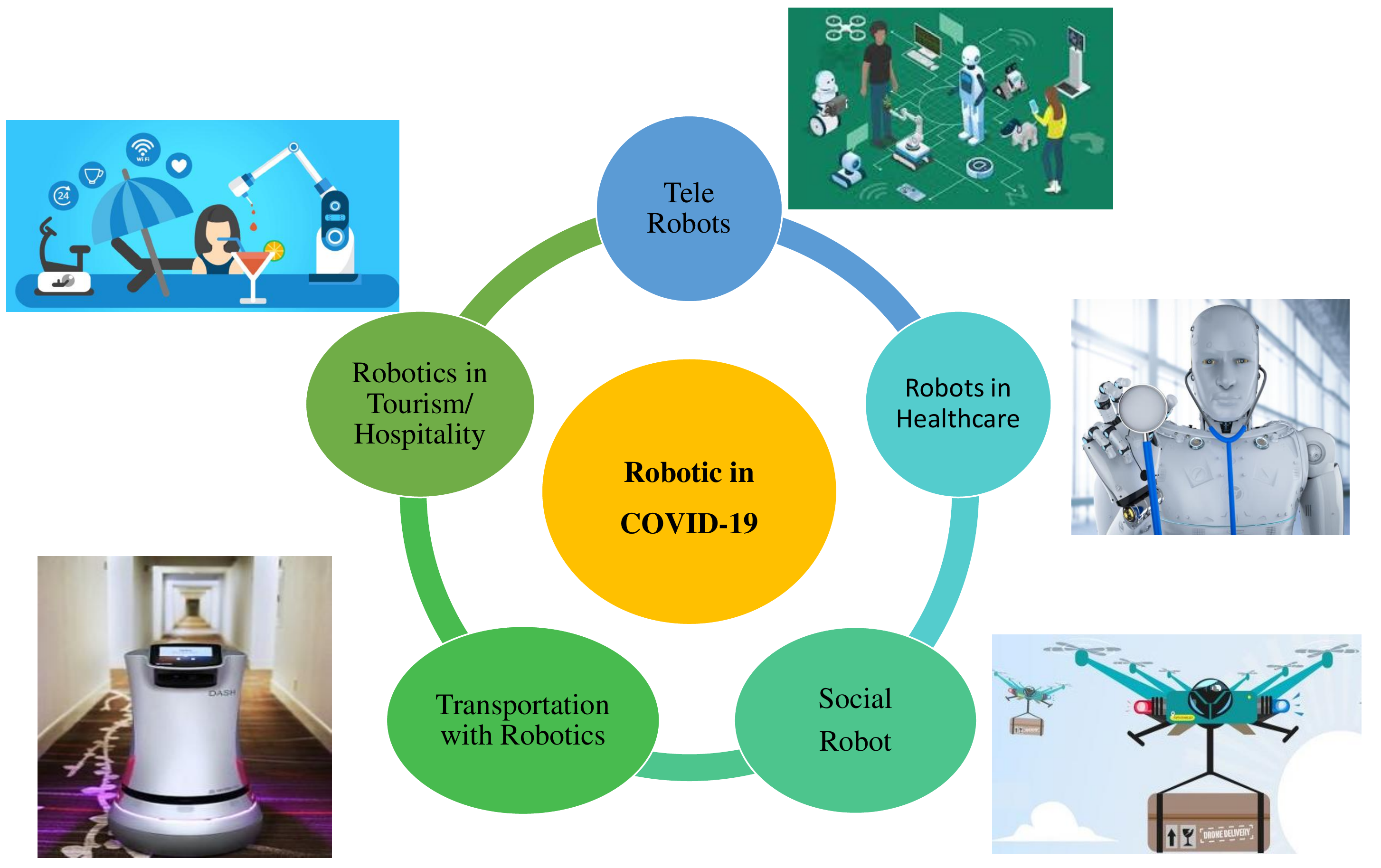}
	\caption{Role of Robotic during pandemic outbreak.}
	\label{FIG:Role_of_Robotics}
\end{figure}

The robotic applications have been used extensively in hospitals, restaurants, transportation, airports, hotels, in various ways to ensure the minimum human contact to avoid the spread of COVID-19 (see Fig. \ref{FIG:Role_of_Robotics}).  A simple example, even robots serving as waiters in the restaurants. The autonomous vehicles, drones and intelligent robots help in sterilising public area, delivering materials, the measurement of body temperature, collection of samples from corona virus patients, providing value added service to patients, and also looking towards security or safety aspects. The advancement of technology and various intelligence techniques have helped in tourism and hospitality management. They are useful in providing service such as food delivery, housekeeping, concierge service and other related tasks during this pandemic situation. Tele robot help in sensing environment and consequently taking necessary reactive action automatically through various machine and deep learning algorithms. They are useful in placing and picking the things, delivering and cleaning actions. Teleoperator type of robots can be very useful in such epidemic situation where tasks can be performed with robotic control mechanism from human. Social robot aid in social activities with interacting human in an acceptable manner \cite{aymerich2020implementation}. They can also assist in healthcare, entertainment industry, teaching and providing comfort communal services. If the growth of robotic based application continues with same pace then it has been predicted that approximately 600 millions jobs around the world and one quarter hospital staff in the USA would be replaced with robots \cite{zeng2020high}. The role of robot is very crucial in tourism and hospitality during this crisis situation to improve the quality of service, safety measures and expectations. The embracing of robotic technology in such situation where social distancing from the human is desired is welcomed. The self-belief in such technology from the people would gradually grow with more trust and minimal risk. The spread of COVID-19 has shown growth of robotic application all over the world. Their role in various field is expected to grow in many areas to improve the sustainability and quality of life.

\subsection{Role of Industry 4.0}

\begin{figure}[htbp]
	\centering
		\includegraphics[width=0.98\textwidth]{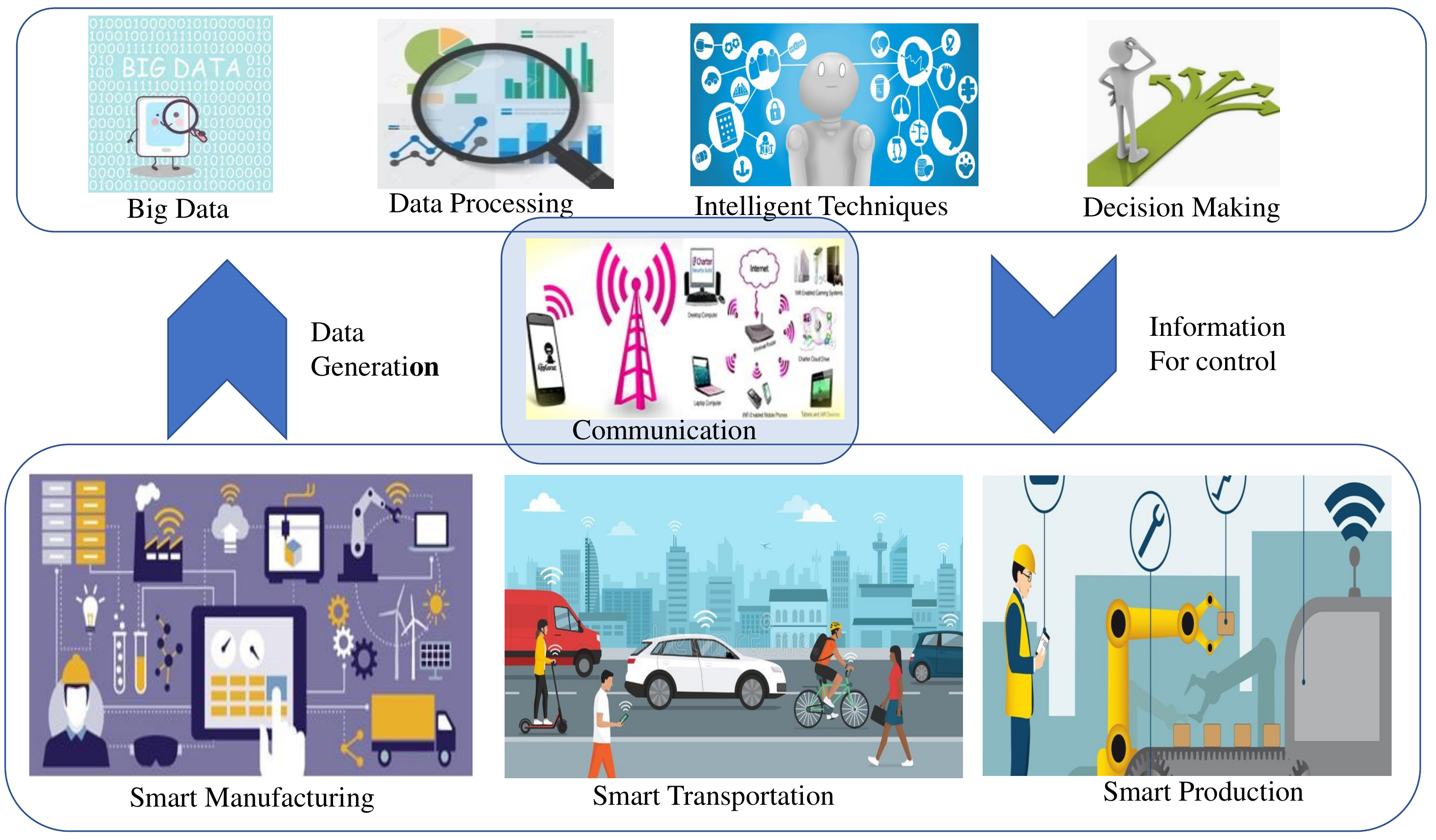}
	\caption{Role of Industry 4.0 during pandemic outbreak.}
	\label{FIG:Role_of_Industry4-0}
\end{figure}

The outburst of COVID-19 has evolved the demand of necessary healthcare equipment (insulin pump, glucometer) for diabetes patient who are at higher risk than others (see Fig. \ref{FIG:Role_of_Industry4-0}). The fourth generation industry revolution industry 4.0 has potential to cater the demand with advanced digital technology. It would be beneficial to provide smart system where real-time information would be processed through Artificial Intelligence (AI), machine learning (ML), intelligent data analytic techniques and Industrial Internet of Things (IIoT) \cite{Mohanty_IEEE-MCE_2020-Jul_Editorial}. The automation of technology has been possible with aid smart manufacturing. The development of any medical device/system could be done rapidly with advances in manufacturing technologies such as 3D printing. The industry 4.0 has made possibility of connected environment where all the technologies are fasten to exchange information for the development of vaccine, healthcare facility, surveillance systems and necessary measures without much human involvement. Industry 4.0 could detect and predict the prevalence of COVID-19 with acquiring data from the smart systems. It can also allow to tackle the diabetes management in better way through intelligent technological solutions. The security solution with block chain technology, through cryptography approaches for peer-to-peer system has resolved the security concerns at various physical locations with suitable traceable mechanism.

\subsection{Trusted Food Supply Chain Management}

Food supply chain is a part of Internet-of-Agro-Things (IoAT) based Agriculture Cyber-Physical System (A-CPS) that makes smart agriculture \cite{Udutalapally_arXiv_2020-May-13-2005-06342}. However, food or diet is important part of smart healthcare and needs discussion.
The consumption of counterfeit and adulterated foods can also affect the immunity of person and may act as a gateway of COVID-19 \cite{naja2020nutrition}. The good quality of food is really important for the consumers and required some technology to identify them easily. This is very crucial in this pandemic situation where healthy diet could always help to maintain good immunity against any kind of virus infection. There are number of food product of fraudulent category  which are available in order to get some financial benefits. It has also been observed that the food contents of packaged food may not be the similar as it mention. While in some cases the product itself is counterfeit. In present scenario, people are trying to consume immune boosting product to maintain their strong immunity but quality of the same product is compromised. Therefore, it is required to develop such kind of technology that can identify a safe and authentic food product.  It is really challenging task to find such kind of cost-effective solution which will not increase the cost of product.

It is essential to develop portable and economical  device to provide the quality assurance of food throughout its life cycle from the production to final consumption stage. This kind of solution would be not only useful in such pandemic situation but could be useful all the time to have proper diet management. This device should be compact and user friendly product at consumer end. The solution is also required to be robust in order to its usage at various stages of the supply chain. There is not such type of solution is available in the market till date, however few researchers have attempted with Surface Enhanced Raman scattering (SERS). The days are not far where such type of solutions would be available in the market.

\begin{figure*}[htbp]
	\centering
	\includegraphics[width=0.90\textwidth]{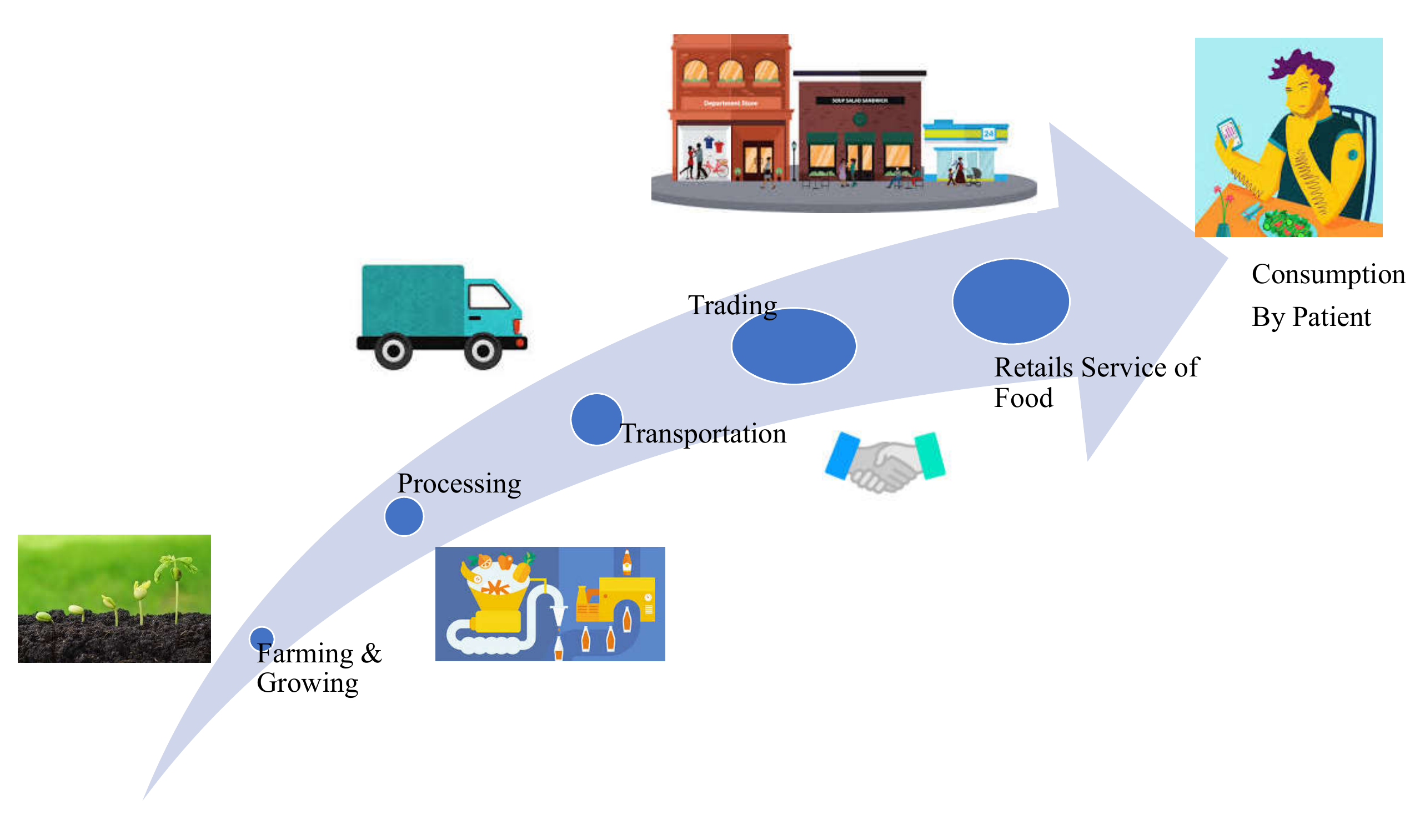}
	\caption{A Vision of Trusted Food Supply Chain: from Farm to the Dinning Table.}
	\label{FIG:Trusted_Food_Supply_Chain}
\end{figure*}


The blockchain has a potential to make great impact in the food supply chain for smart agriculture \cite{Rachakonda_arXiv_2020-Jul-2007-07377_SaYoPillow, Alkhodair_ISVLSI_2020_McPoRA, kohler2020technology}. The technology could be useful to analyse the quality of the food product through the reliable supply chain management framework.  The sustainable system  would be developed using block chain technology which is helpful to develop trust and transparency from its production to consumption. This would be beneficial for diabetes patients to maintain their proper healthy diet plan during this COVID19 outbreak.

\section{Smart Healthcare for Diabetes during Pandemic Outbreak}
\label{Sec:Smart_Healthcare_for_Diabetes}

There is some prediction that the novel corona virus SARS-CoV-2 may not go from the world in near future. Therefore, each and individual has to change their life style and take all the preventive actions against the virus. A diabetes patients are more susceptible, they have to adopt some healthcare measure in their daily life. There are few smart healthcare solutions which are  useful during this COVID-19 crisis is discussed in the current Section.

\subsection{Telemedicine for Diabetes Patients}

The COVID-19 pandemic has made the impact drastically on healthcare organizations all over the globe. Consequences of this lockdown for diabetic patients would result in  reduced values of insulin and antihyperglycemic components. The mandate on social distancing 
has also posed the restrictions on routine visits for the patients to doctors. The role of telemedicine is significant during such unparalleled situation which allow patient to manage their health profile for the diabetes alike chronic diseases \cite{joshi2020diabetes}.

\begin{figure}[htbp]
	\centering
	\includegraphics[width=0.95\textwidth]{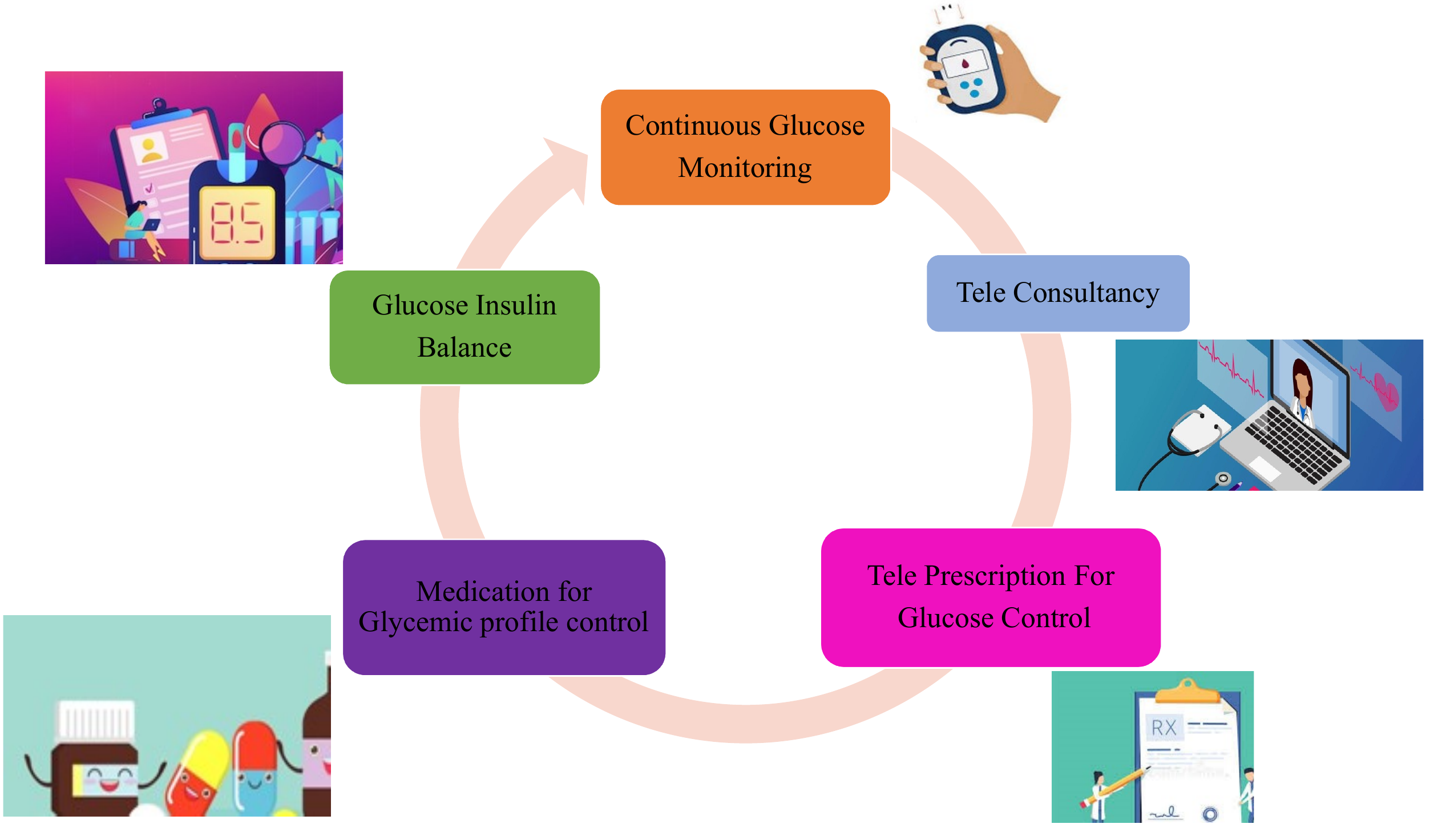}
	\caption{Role of Teleedicine during Pandemic Outbreak.}
	\label{FIG:Role_of_Telemedicine}
\end{figure}

The telemedicine service could be classified as per the time frame and mode of
communication with health service provider (See Fig. \ref{FIG:Role_of_Telemedicine}).  The telemedicine could be provided through text emails, Fax, short messaging service, emails, Fax and conversion by chat on social network platforms. There are several video conferencing software. There are few traditional way of taking proper medication such as voice over internet protocol, phone etc. This kind of telemedicine service would be important for situations where less numbers of doctors per person available and would be able to reach rural area where it is scarcity of the health service. The service of telemdicine would be also popular gradually with the increase of smart-phones day by day. This kind of service for diabetes patients is considered as blessing where the vulnerability gets reduced from COVID-19 infection.

Through this any physician can connect with the patient and give him advice by examining his various medical diagnosis like prevailing history, self-monitored blood glucose charts (SMBG) and self-monitored blood pressure (SMBP) values. Tele medicine has come a long way and its usage guided diabetic patients. A small study, where 35 randomised controlled trials
(RCTs) done with aid of  telemedicine (like video, phone and email) in China (where a pool of (n,3514) population) was given the consultancy for 36 months. This experimentation demonstrated decrease in HbA1c by 0.37\% (p < 0.001) in group telemedicine as compared to other. Similar review was conducted by Flodgren and colleagues,this was done on 21 RCTs of patient with diabetes (n,2768). They were given an interactive session through (real time video or remote monitoring) in an appendage to only standard care alone. With this measure a overall decrease of HbA1c by 0.31\% (p < 0.001) in patients on telemedicine as set side by side with controls. A review article recently published showed by 46 studies including both the types of diabetes with patients of type-2 diabetes mellitus (T2DM, $n$, 24000) and type-1 diabetes mellitus (T1DM $n$, 2052) were experimented with different modes of telemedicine. This investigation reported an overall reduction in mean value of HbA1c by T1DM (0.12 to 0.86\%) and T2DM (0.01\% to 1.13\%) patients, respectively. 

India is in its early stage of research on telemedicine and diabetes. The Pilot project known as ``Diabetes Rath'' (Hindi name of Mobile Vehicle) was started with aiming of spreading knowledge of diabetes care using mobile van facilitated with telemedicine (like computer and video conferencing app for transmitting retinal images to ophthalmologist, consult with diabetes foot specialist and diabetologist at a tertiary care centre ) especially in underprivileged areas \cite{gopalan2019diabetes}. This study has shown positive results in screening and managing diabetes through telemedicine.

There are several general guidelines for telemedicine service as follows:
\begin{itemize}
	\item The confidential and private information should be 
	maintained properly.
	\item The personal information related to patient such as name,age, address  should be maintained. 
	\item The medical history and records for the patient medical records should be kept with their prescription along with the any testing records.
	\item The consultancy charge of the patient could be decided as per his/her medical prescription and the procedure of getting tele-consultancy. 
\end{itemize}

Telemedicine would allow the possibility to expand the medical facility and healthcare sector every part of the world without any kind of geographic barrier. Consequently, telemedicine platform can help to reach the health related services to places where there is any constraint of physical structures. It would gain massive popularity in the current epidemic where social distancing is the  need of the society. The remote consulting with doctor is really attractive solution for urban tertiary where medical centres faces real challenge to accommodate each patient with proper point of care facilities. Telemedicine option is also useful when it is required to coordinate complex multidisciplinary care via a tumour board conference format among specialists that are geographically separated.

\subsection{Non-Invasive measurement glucose measurement and automatic glucose control}

\begin{figure}[htbp]
	\centering
	\includegraphics[width=0.75\textwidth]{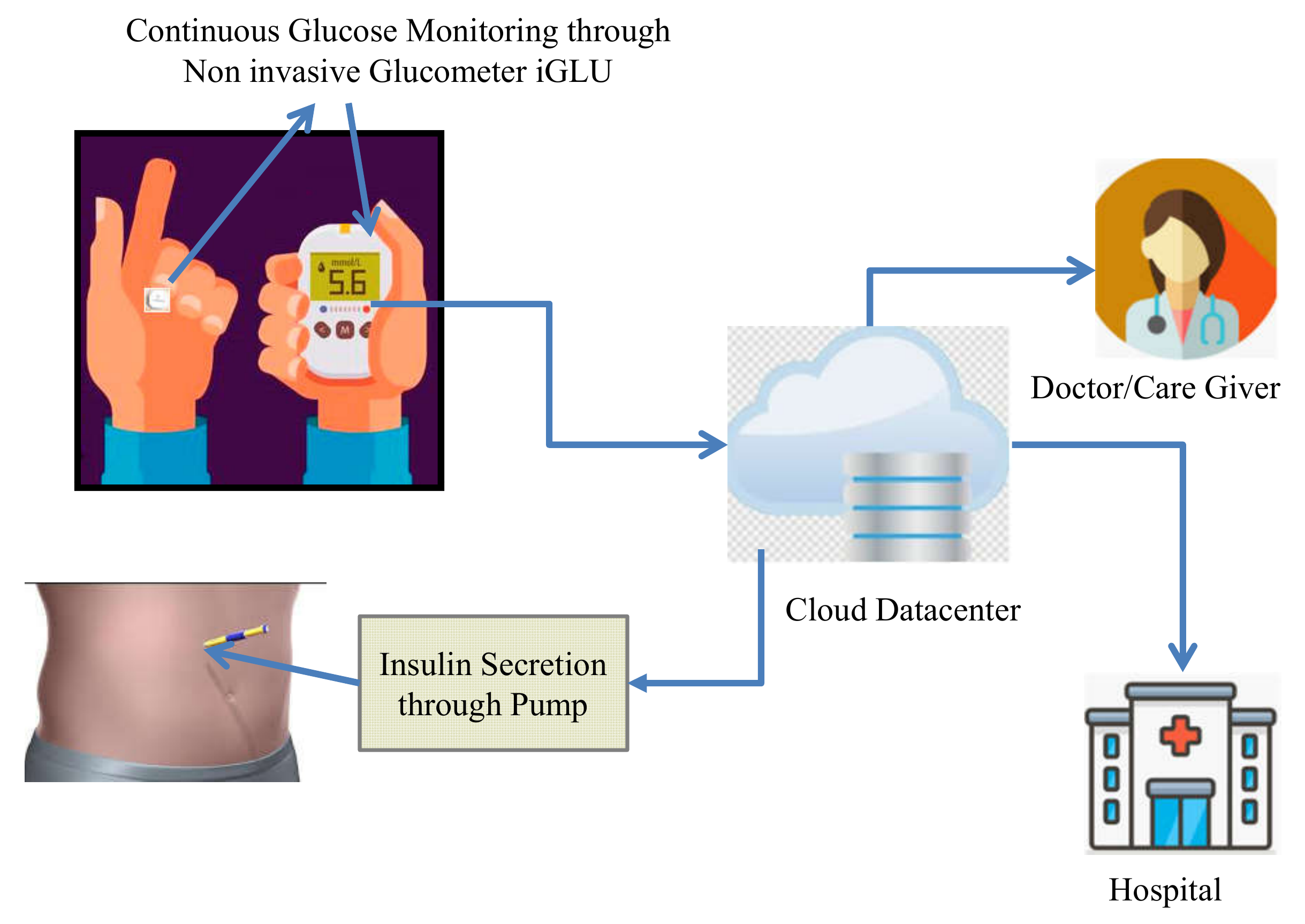}
	\caption{Our Intelligent Non-Invasive Glucose Monitoring with Insulin Control Device (iGLU) \cite{Jain_IEEE-MCE_2020-Jan_iGLU1, Jain_arXiv_2020-Jan-28-2001-09182_iGLU2, Jain_WF-IoT_2020_iGLU, Joshi_ISVLSI_2020_Secure-iGLU}.}
	\label{FIG:Secure-iGLU_Illustration}
\end{figure}

The non-invasive glucose measurement device (such as iGLU, see Fig. \ref{FIG:Secure-iGLU_Illustration}) would be the state of art solution for frequent glucose measurement \cite{Jain_IEEE-MCE_2020-Jan_iGLU1, Joshi_TCE_2020.3011966_iGLU2, Jain_arXiv_2020-Jan-28-2001-09182_iGLU2, Jain_WF-IoT_2020_iGLU, Joshi_ISVLSI_2020_Secure-iGLU}. The proposed solution will be useful for continues glucose monitoring and its control with insulin secretion along with integration of Internet of Medical Things (IoMT) based healthcare Cyber-Physical System (H-CPS) framework. iGLU would help to provide instant diagnosis to remote located diabetes patient through telemedicine in this crisis situation. The conventional blood picking process would not be an ideal choice for frequent measurement for older population and children. It has also associated risk with trauma and touching the needle could be source for infection. For diabetes patients, if the blood sugar goes above 250, then it would be at risk for ketones. The ketones is considered as the poison for body. During this pandemic, it is very vital to keep these ketone levels down. The continues monitoring solution would help to balance the glucose profile in the body. The glucose insulin model has been proposed to analyse the plasma insulin variation  with proper insulin secretion plan. The proposed insulin delivery system is integrated with IoMT to have prescribed diet and insulin plan under the supervision of remotely located doctor. The insulin pump would be helpful to have the proper glucose-insulin level that is important to improve the immune system to avoid COVID-19 infection.

\subsection{Intelligent Diet Control for Glucose Insulin Balance}

\begin{figure*}[htbp]
	\centering
	\includegraphics[width=0.80\textwidth]{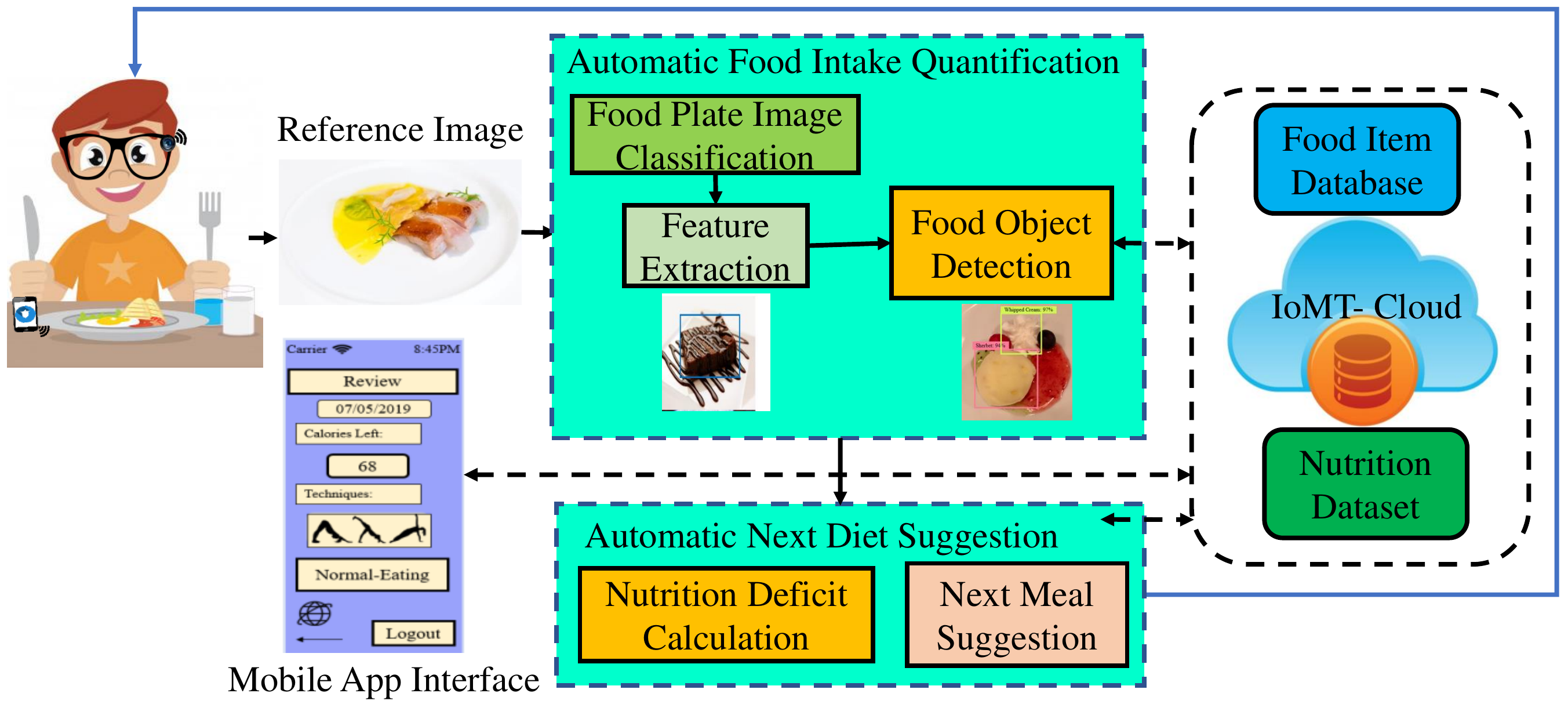}
	\caption{Diet Automatic Monitoring and Control for Blood Glucose Control \cite{Rachakonda_TCE_2020-May, Sundaravadivel_TCE_2018-Aug}.}
	\label{FIG:Automatic_Diet_Monitoring_Control}
\end{figure*}

Unhealthy diet can also cause several health issues. The diabetes patient has to plan their diet to control glycemic profile. Automatic IoMT-based mechanisms which  allow user to monitor their food intake and help to create awareness about the right kind of food with suggestions for next diet (see Fig. \ref{FIG:Automatic_Diet_Monitoring_Control}) can be crucial for diabetic people.  iLog educate users for Normal-Eating and Stress-Eating, where Stress Eating defines the uncontrollable consumption of high caloric foods. It also provide a fully-automated edge level device which is useful to detect automatic stress variation in healthy lifestyle. The iLog framework uses a mobile platform as a user interface. The images of  foods are being detected automatically and subsequently  quantified foods are compared with the images of stored database \cite{Rachakonda_TCE_2020-May, Sundaravadivel_TCE_2018-Aug}. The proposed approach could be helpful to have proper glucose-insulin balance for diabetes patients which ultimately helps to minimize the chances of COVID-19 infection. It provides a fully-automated platform for healthy life style with monitoring the stress behavioural by virtue of no input  from user end.

\subsection{Rapid Detection of COVID-19}

\begin{figure}[htbp]
	\centering
	\includegraphics[width=0.55\textwidth]{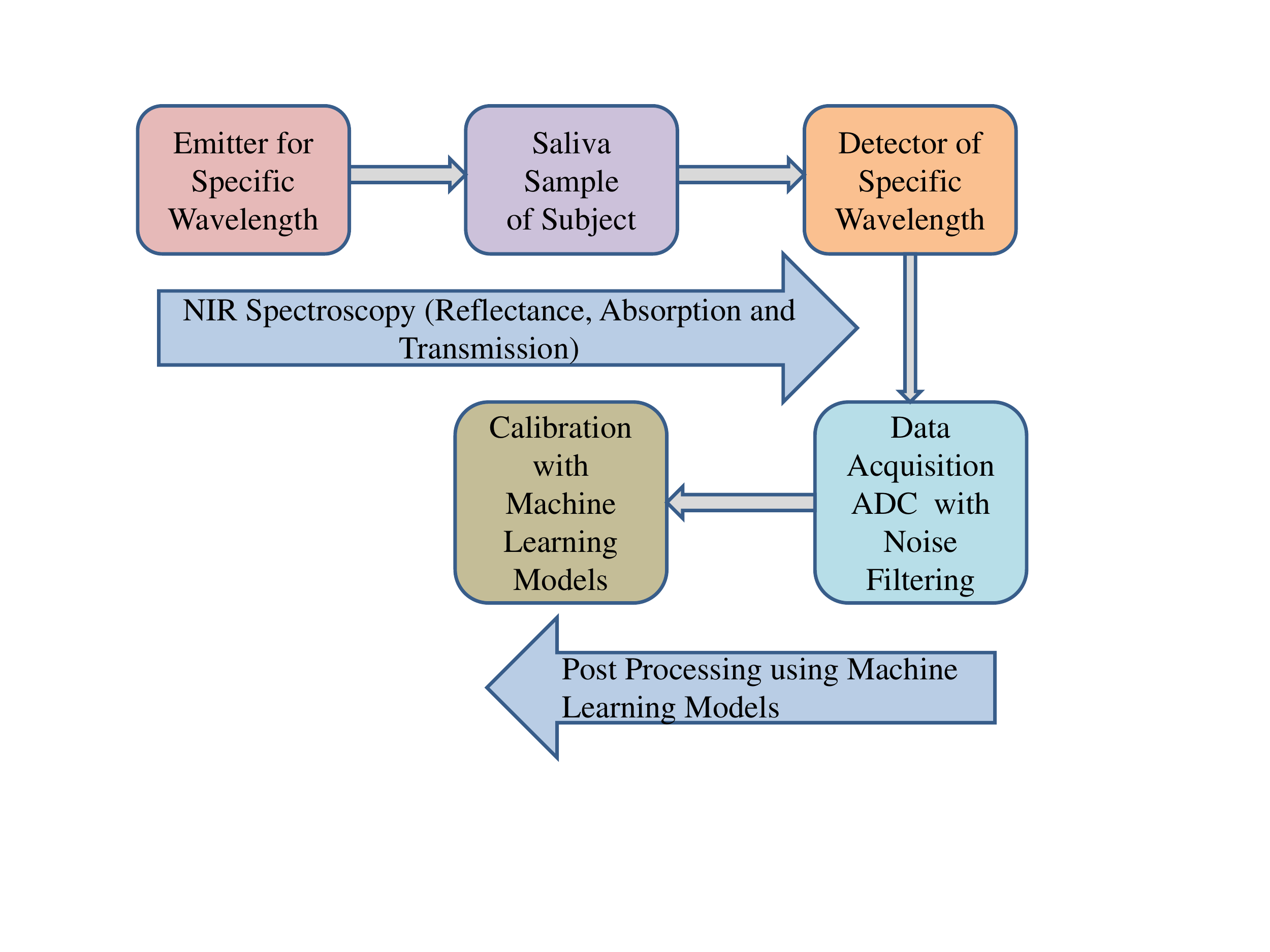}
	\caption{Rapid Detection of COVID-19 using NIR Spectroscopy.}
	\label{FIG:Rapid_Detection_COVID-19}
\end{figure}

The development of rapid, accurate and portable  diagnostic technique for corona virus is also required for the diabetes patients. If complications of  diabetic patients increase then they are in more danger. The near infrared (NIR) spectroscopy could be the breakthrough for this purpose. The focus is to develop a system for the Novel Coronavirus (COVID-19) detection using optical technique (see Fig. \ref{FIG:Rapid_Detection_COVID-19}). The proposed system shall use light with specific wavelengths for instant measurement. The proposed technology would help to sense the novel coronavirus from saliva  of the patient. The proposed system is on chip (SoC) with specific emitters and detectors, analog-to-digital converter (ADC) and acquisition module be embedded to process the sample through saliva. The acquired values would be then processed through machine learning models to detect the presence of corona virus. The obtained values require to be calibrated with standard reference value. The novel system would have higher precision level and stability in measurement compared to traditional measurement. The system would be capable to test patients within few minutes.  The system is to be integrated with an IoT framework for data storage where remote instant diagnosis is possible through shared server data of the patient. The proposed solution shall be cost-effective and provide instant measurement of COVID-19 infection. The similar technology was applied successfully for rapid detection of Zika virus and hepatitis B and C virus infection  \cite{roy2019spectroscopy, fernandes2018rapid}.

\subsection{Wearable Safety-Aware Mobility Tracking Device}

\begin{figure}[htbp]
	\centering
	\includegraphics[width=0.75\textwidth]{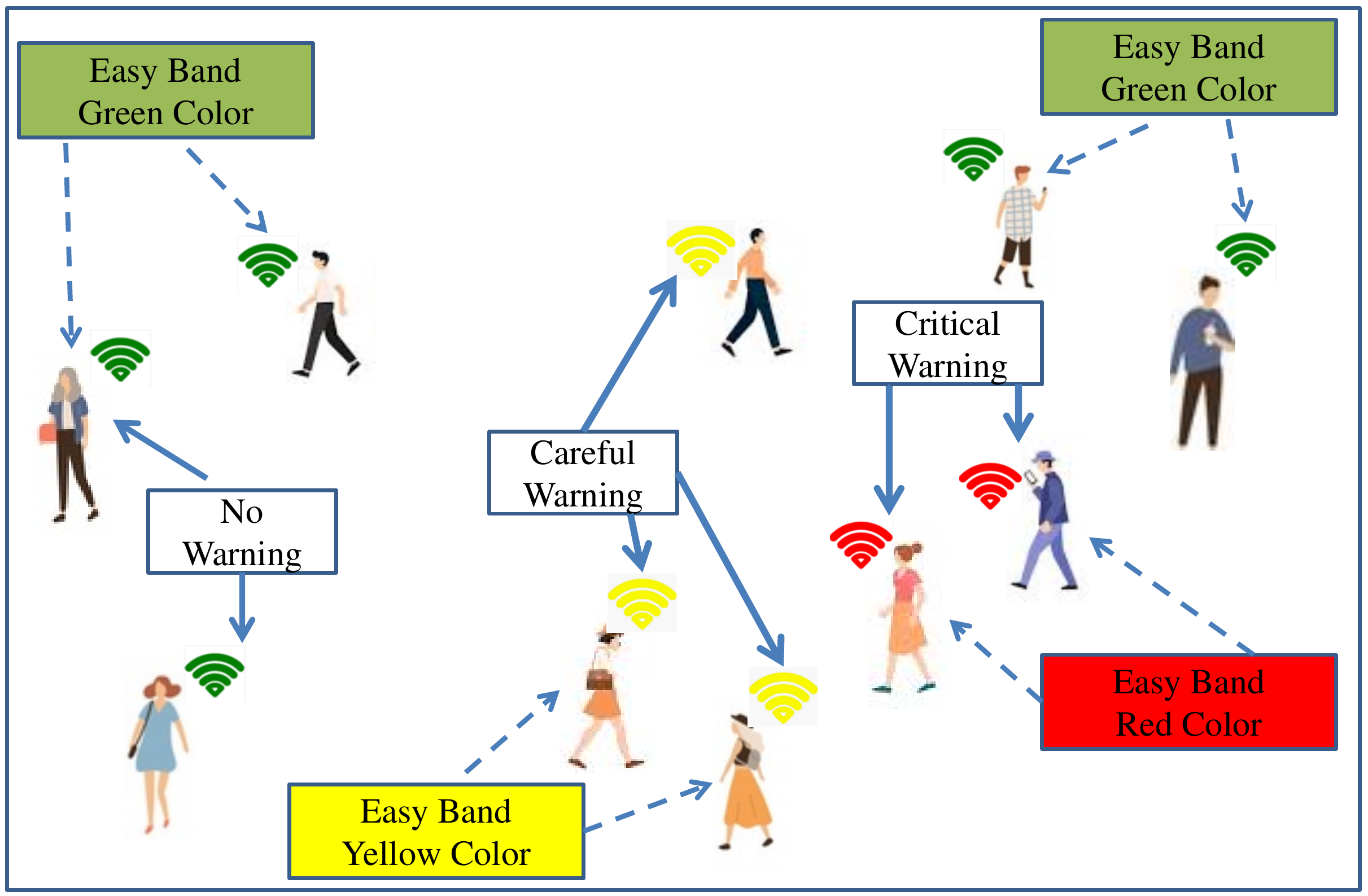}
	\caption{Wearable wristband for contact tracing \cite{Tripathy_MCE_2020-EasyBand}.}
	\label{FIG:EasyBand_Social_Distancing_Theme}
\end{figure}

An IoMT device easy-band based sensors is useful to sense the presence of another COVID-19  patient within a radius of 6 to 13 feet \cite{Michael_MCE_2020_3002492} (See Fig. \ref{FIG:EasyBand_Social_Distancing_Theme}). It is wearable, accurate and cost effective solution of monitoring COVID-19 patient and would be integrated with IoMT/H-CPS framework. The diabetic patients are of great need of such type of device which would help to track their health in this COVID-19 outbreak.

All must have to wear this device for safety aware mobility tracking purpose. The device comprises of three LED colors mainly as red (highly suspected), Green (safe) and yellow (mildly suspected) \cite{Tripathy_MCE_2020-EasyBand}. Every device is capable of storing the local information such as timestamp, time period and device ID of those devices which are in contact of the present device within area of 6 feet zone. The device is able to store these information upto last 14 days. The healthy tested persons (with diabetes or pre-diabetes) would have the access of this device as mobility pass with active green light. The person has to wear the device once having the possession of this device and will not attempt for the removal. In case of such any efforts of tempering, it will be directly reported  to concerned authority. If any person comes in close contact of another COVID-19 person then the device comes at red light state and record the details of patient. The information are updated on server at regular basis. Device automatically starts vibrating to alert when it come in an area of 4 meters (13 feet) in vicinity of  yellow or red device. Another alert is generated by a beep sound (critical warning) when it comes in contact with yellow or red device. When a green device spends a long time in close contact with a yellow/red device, its status will automatically change to yellow. This device will also have a temperature sensor (e.g. infrared/IR sensor) to sense the body temperature and measure respiratory of the person for issuing necessary preventive actions and changes status from green to yellow. This way the device will help the diabetic patients (or any citizen) to stay safe by automatically sensing the suspects. The battery life of the devices has been enhanced with ultra low power operation. To social distance while traveling, one has to sense the colour that pop up on the device which can warns people through their signals.

With the pandemic around, quarantining a potential carrier of  COVID-19 can curb the spread to a certain extend. However, with the difficulty in pinpointing a carrier of virus many countries are adhering to other measures like shelter-in-place, stay-at-home, and lockdown. Prolonged stay-at-home measure has created many other problem such as economical crises, unemployment, food scarcity, and mental health problems of individuals etc. Again contact tracing of positive cases and isolating them will be a hectic, unreliable and error prune task. As a result again re-imposition of lockdown is essential. Nevertheless, if the mentioned solution is controlled with a technological approach, it would help in stabilizing the current scenario. SARS-CoV2 effects more to elderly people, children and person with pre-existing diseases, such pre-diabetes, and diabetes. Such type of safety aware device would be really helpful to diabetic community for their self-care measurement.

\section{Conclusion and Future Directions}
\label{Sec:Conclusion}

The article attempts to spread awareness among the people (specially diabetes patients) with possible recommendation and future technologies for smart healthcare. 
The case studies around the globe suggest that diabetic people are more vulnerable to COVID-19 infection and also suffer from more severity in terms of medical complications. The underneath diabetic patients have higher risks of two to three times compared to non-diabetic person.
The studies show that a better control over glycemia can be advantageous to the patient simultaneously suffering from both diabetes and viral respiratory diseases such as COVID-19.
The proper diabetic care of the person would help in reducing the prevalence of  COVID-19
infection. The balance glucose insulin profile could improve the morbidity rate as well as mortality rate against SARS-CoV-2. It has potentials to reduce length of stay in the infected COVID-19 patients and also useful in avoiding the wide spared of virus among the community. We have discussed many devices and techniques available for diet management, stress management, and glucose-level management which can be helpful.


Presently, significant effort going on around the globe for the development of vaccine  and medicines against novel corona virus COVID-19 and many agencies across the world are working  for the success of the same. However, there is requirement to have proper planning so it have early access of most vulnerable people such as diabetic, cardiovascular disease, hyper tension people.   
Although there has been substantial development of technological solution, there is a need for smart long term strategy that would require to fight against such pandemic in future in order to minimize the social and economical impact. In this respect, IoMT driven healthcare Cyber-Physical System (H-CPS) will play a major role \cite{Mohanty_IEEE-MCE_2020-Sep_Editorial}.


\bibliographystyle{IEEEtran}


\section*{Authors' Biographies}

\begin{minipage}[htbp]{\columnwidth}
\begin{wrapfigure}{l}{1.3in}
\vspace{-0.4cm}
\includegraphics[width=1.3in,keepaspectratio]{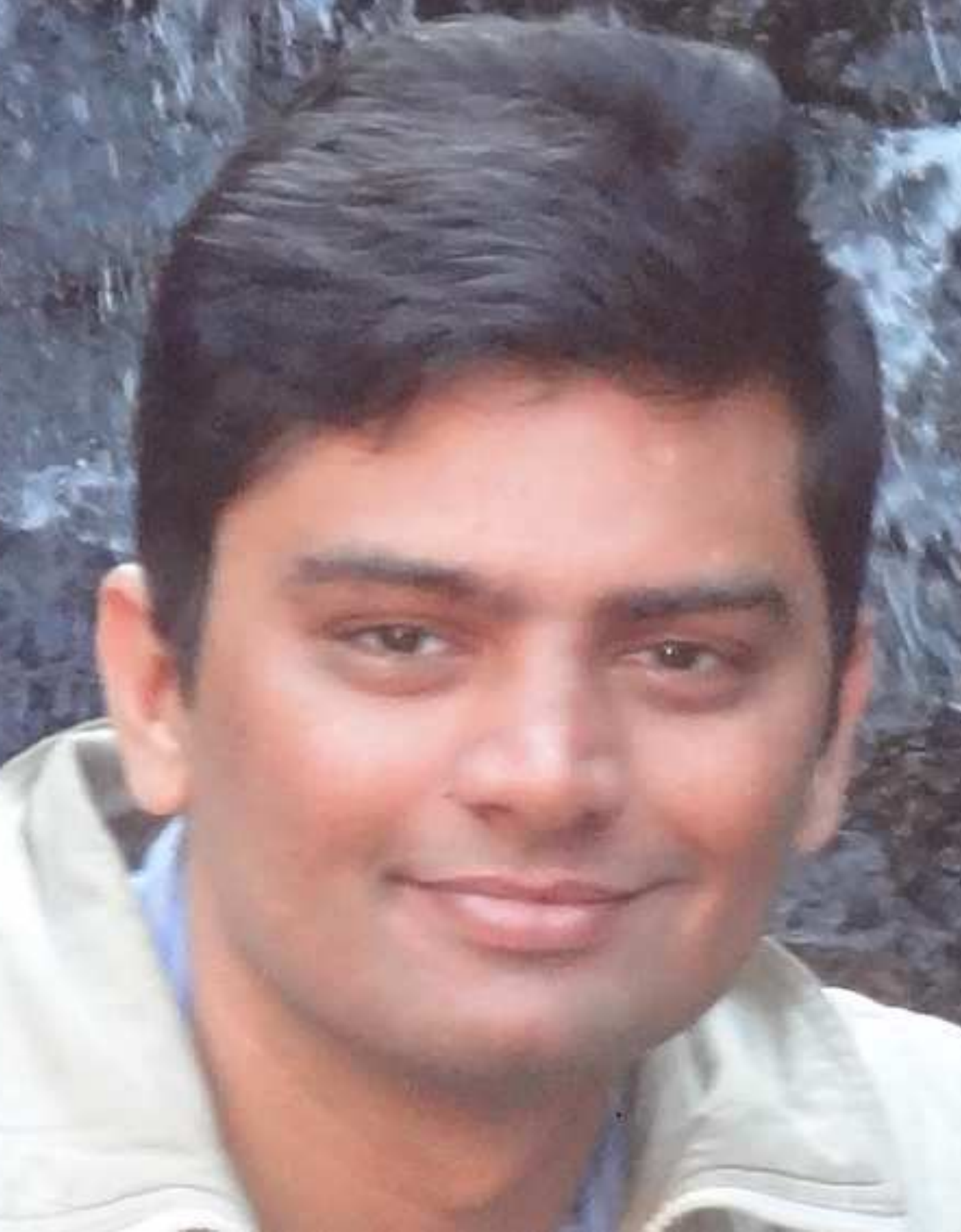}
	\vspace{-0.5cm}
\end{wrapfigure}
\noindent
\textbf{Amit M. Joshi} (M'08) has completed his M.Tech (by research) in 2009 and obtained Doctoral of Philosophy degree (Ph.D) from National Institute of Technology, Surat in August 2015. He is currently an Assistant Professor at National Institute of Technology, Jaipur since July 2013. His area of specialization is Biomedical signal processing, Smart healthcare, VLSI DSP Systems and embedded system design. He has published six book chapters and also published 50+ research articles in peer reviewed international journals/conferences. He has served as a reviewer of technical journals such as IEEE Transactions, Springer, Elsevier and also served as Technical Programme Committee member for IEEE conferences. He also received UGC Travel fellowship, SERB DST Travel grant  and CSIR Travel fellowship to attend IEEE Conferences in VLSI and Embedded System. He has served session chair at various IEEE Conferences like TENCON -2016, iSES-2018, ICCIC-14. He has supervised 18 M.Tech projects and 14 B.Tech projects in the field of VLSI and Embedded Systems and VLSI DSP systems. He is currently supervising six  Ph.D. students.
\end{minipage}

\vspace{1.8cm}

\begin{minipage}[htbp]{\columnwidth}
	\begin{wrapfigure}{l}{1.3in}
		\vspace{-0.4cm}
		\includegraphics[width=1.3in,keepaspectratio]{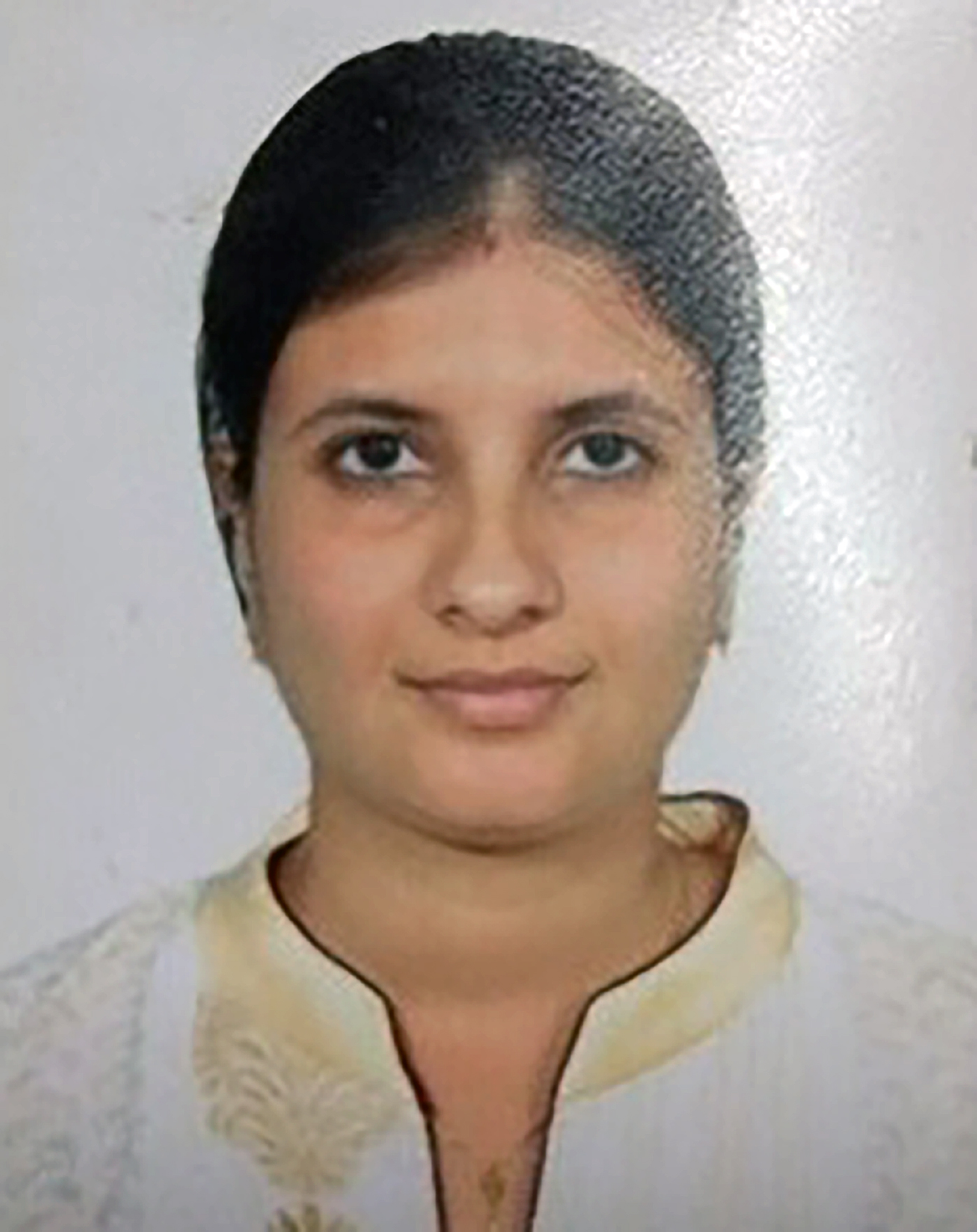}
		\vspace{-0.5cm}
	\end{wrapfigure}
	\noindent
	\textbf{Urvashi P. Shukla} received her bachelor's degree in Electronics \& Communication Engineering from C. K . Pithawala College of Engineering and Technology, affiliated under Veer Narmad South Gujarat University, Gujarat, India. She pursued her M.E from Sarvajanik College of Engineering and Technology, Surat, Gujarat, India under Gujarat Technological University. She completed her Ph.D from NIT, Jaipur, Rajasthan, India. She has published papers in reputed journals and conferences. Her research interests revolves to obtain solution for problems associated with hyperspectral Images by incorporating various optimization algorithms based on heuristic approach.
\end{minipage}

\vspace{1.8cm}

\begin{minipage}[htbp]{\columnwidth}
	\begin{wrapfigure}{l}{1.3in}
		\vspace{-0.3cm}
		\includegraphics[width=1.3in,keepaspectratio]{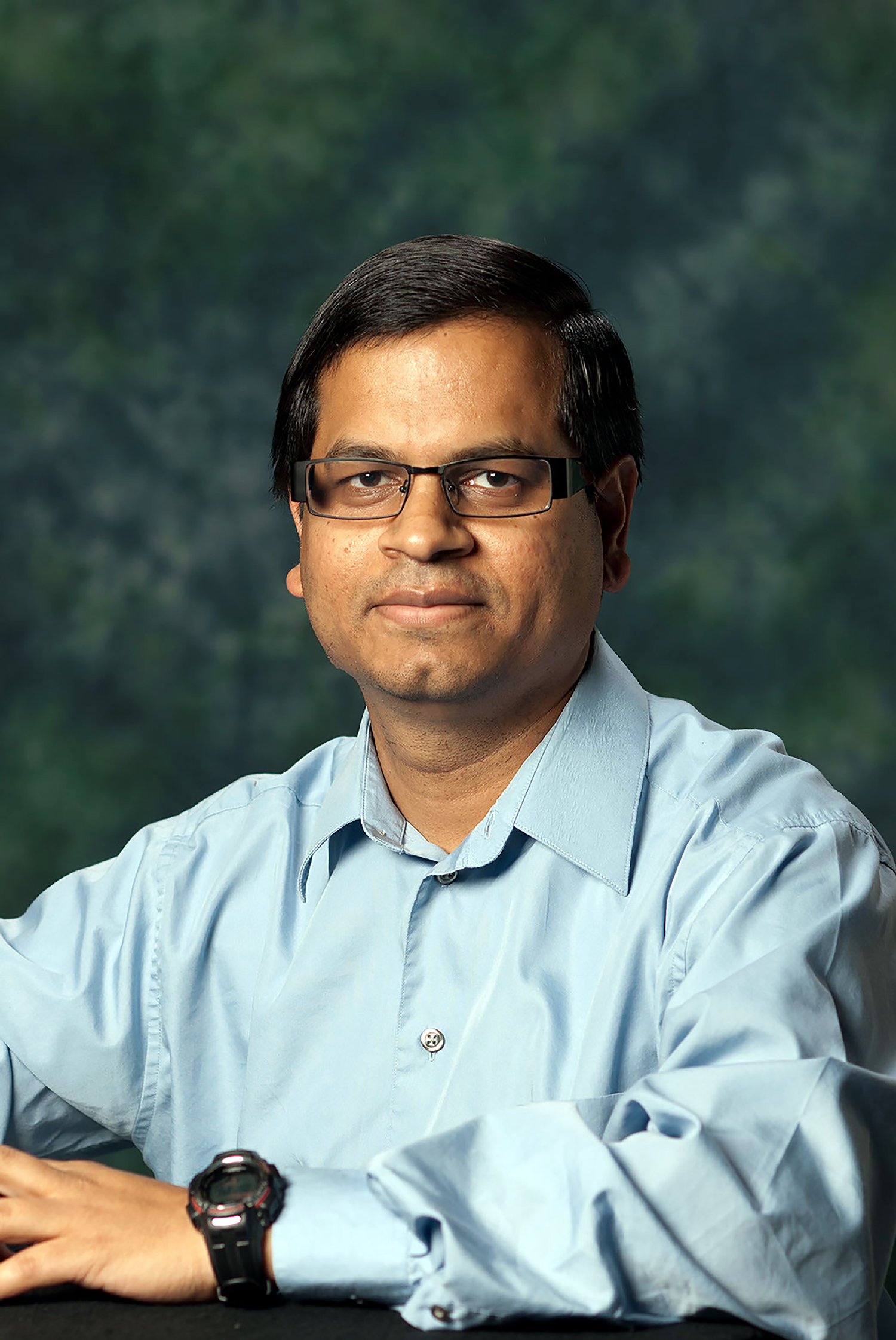}
		\vspace{-0.5cm}
	\end{wrapfigure}
	\noindent
\textbf{Saraju P. Mohanty} (SM'08) received the bachelor's degree (Honors) in electrical engineering from the Orissa University of Agriculture and Technology, Bhubaneswar, in 1995, the master's degree in Systems Science and Automation from the Indian Institute of Science, Bengaluru, in 1999, and the Ph.D. degree in Computer Science and Engineering from the University of South Florida, Tampa, in 2003. He is a Professor with the University of North Texas. His research is in ``Smart Electronic Systems'' which has been funded by National Science Foundations (NSF), Semiconductor Research Corporation (SRC), U.S. Air Force, IUSSTF, and Mission Innovation. He has authored 350 research articles, 4 books, and invented 4 U.S. patents. His Google Scholar h-index is 36 and i10-index is 134 with 5800+ citations. 
He introduced the Secure Digital Camera (SDC) in 2004 with built-in security features designed using Hardware-Assisted Security (HAS) or Security by Design (SbD) principle. He is widely credited as the designer for the first digital watermarking chip in 2004 and first the low-power digital watermarking chip in 2006.
He is a recipient of 12 best paper awards, Fulbright Specialist Award in 2020, IEEE Consumer Electronics Society Outstanding Service Award in 2020, the IEEE-CS-TCVLSI Distinguished Leadership Award in 2018, and the PROSE Award for Best Textbook in Physical Sciences and Mathematics category in 2016. He has delivered 9 keynotes and served on 5 panels at various International Conferences. 
He has been serving on the editorial board of several peer-reviewed international journals, including IEEE Transactions on Consumer Electronics (TCE), and IEEE Transactions on Big Data (TBD). 
He is the Editor-in-Chief (EiC) of the IEEE Consumer Electronics Magazine (MCE). 
He has been serving on the Board of Governors (BoG) of the IEEE Consumer Electronics Society, and has served as the Chair of Technical Committee on Very Large Scale Integration (TCVLSI), IEEE Computer Society (IEEE-CS) during 2014-2018. 
He is the founding steering committee chair for the IEEE International Symposium on Smart Electronic Systems (iSES), steering committee vice-chair of the IEEE-CS Symposium on VLSI (ISVLSI), and steering committee vice-chair of the OITS International Conference on Information Technology (ICIT). 
He has mentored 2 post-doctoral researchers, and supervised 12 Ph.D. dissertations, 26 M.S. theses, and 10 undergraduate projects.
\end{minipage}


\end{document}